\documentclass[10pt,a4paper,aps,pra,twocolumn,showpacs,superscriptaddress]{revtex4-1} 
\usepackage[english]{babel}

\usepackage{url} 
\usepackage{hyperref}
\usepackage{amsmath}
\usepackage{slashed} 
\usepackage{graphicx}
\usepackage{color}
\usepackage{bm} 
\usepackage{mathtools,mathrsfs}
\usepackage{amsmath,amssymb}
\usepackage{bbold}

\usepackage{bm,bbm}
\usepackage{siunitx}

\usepackage[normalem]{ulem}

\usepackage{cleveref}

\begin{document}
\allowdisplaybreaks

\title{Preprint version of: GPU Accelerated Simulation of Channeling Radiation of Relativistic Particles}
\author{C. F. Nielsen}
\email{christianfn@phys.au.dk}
\affiliation{Department of Physics and Astronomy, Aarhus University, 8000 Aarhus, Denmark}
\date{\today}
\begin{abstract}
In this paper we describe and demonstrate a C++ code written to determine the trajectory of particles traversing oriented single crystals and a CUDA code written to evaluate the radiation spectra from charged particles with arbitrary trajectories. The CUDA/C++ code can evaluate both classical and quantum mechanical radiation spectra for spin 0 and 1/2 particles. We include multiple Coulomb scattering and energy loss due to radiation emission which produces radiation spectra in agreement with experimental spectra for both positrons and electrons. We also demonstrate how GPUs can be used to speed up calculations by several orders of magnitude. This will allow research groups with limited funding or sparse access to super computers to do numerical calculations as if it were a super computer. We show that one Titan V GPU can replace up to 100 Xeon 36 core CPUs running in parallel. We also show that choosing a GPU for a specific job will have great impact on the performance, as some GPUs have better double precision performance than others.

\bigskip
\end{abstract}

\maketitle

\section{Introduction}
The problem of evaluating the radiation spectrum from accelerating charges is relevant in various contexts, and has in many cases been solved analytically \citep{jacksons}. However valuating radiation from arbitrary motion is not always possible by analytical solutions. In this paper we describe a numerical code designed to evaluate both classical and semi-classical radiation from classical particle trajectories. This problem is numerically very difficult since evaluating the radiation spectra involve integration of rapidly oscillating functions. Our code utilizes the enormous parallelism available on GPUs. Thus evaluating radiation from ultra-relativistic particles is now possible with the enhanced double precision capabilities in modern GPUs such as the Nvidia GP100 \cite{gp100}, GV100 \cite{gv100} and the Titan V (which is equipped with the same chip as the GV100 \citep{turing}). 
The code is written in C++/CUDA\footnote{CUDA is a C/C++ parallel computing platform designed to implement the use of GPUs for general purpose programming. This allows the user to launch thousands of parallel processes.
https://www.geforce.com/hardware/technology/cuda}, where the serial task of solving the equation of motion of single particles is done on a CPU, while the radiation integrals are solved on the GPU.

Other codes for the purpose of evaluating radiation from relativistic particles have been developed, see e.g. \cite{MBNexp}, an implementation in the existing Meso Bio Nano explorer software (MBN Explorer). This software is very capable, but for larger calculations it requires large CPU clusters. For many research-groups, gaining access to large CPU clusters is not possible, and we show in this paper that by using GPUs, these very demanding numerical tasks can be performed on personal workstations.

This paper is organized as follows: First we describe the theory which is implemented in the code, then the implementation of the theory described and then we describe several tests against theory and experiment to validate the different implementations. 
Last we compare computation times of the radiation integrals using different CPUs and GPUs.

\section{Radiation integrals\label{radiationintegral}}
From the Lienard-Wiechert fields one obtains the classical radiation spectum, which is evaluated by the integral \cite{jacksons}

\begin{equation}
\frac{d^2I}{d\omega d\Omega} = \frac{e^2}{4\pi^2}\left|\int^\infty_{-\infty}\bold{f}(t,\bold{n}) 
e^{i\omega(t-\bold{n}\cdot\bold{x}(t))}dt
\right|^2,
\label{eq:integralrad1}
\end{equation}
\begin{equation}
\bold{f}(t,\bold{n}) = \frac{\bold{n}\times(\bold{n}-\boldsymbol{\beta})\times\dot{\boldsymbol{\beta}}}{(1-\boldsymbol{\beta}\cdot\bold{n})^2},
\label{eq:integralrad2}
\end{equation}
where $e$ is the electric charge, $c$ is the speed of light, $\omega$ is the frequency of the emitted radiation, $\bold{n}=(\sin\vartheta\cos\varphi,\sin\vartheta\sin\varphi,\cos\vartheta)$ is the direction of emission with polar and azimuthal angles $\vartheta$ and $\varphi$ defined relative to an observer, $d\Omega=\sin\vartheta d\vartheta d\varphi$ and $\bold{r}$, $\boldsymbol{\beta} = \bold{v}/c$ and $\boldsymbol{\dot{\beta}} = \bold{\dot{v}}/c$ are the position, velocity and acceleration of the charge at time $t$.
Since radiation emission from relativistic particles often is of interest, we are dealing with small differences between large numbers. To increase the numerical precision, we do a series expansion of these large quantities, as is done in \cite{TobiasUdvikling}, keeping only the leading order terms of the changes to these large numbers. Assuming the particle travels in the $z$-direction, 
one can solve the equation of motion for the values $\delta z(t) = z(t)-\beta_0t$ and $\delta v_z(t) = v_z(t)-\beta_0$ where $\beta_0 \approx 1-\frac{1}{2\gamma^2}$. Since almost all radiation is emitted in the forward direction we also have that $n_z = \sqrt{1-n_x^2-n_y^2} \approx 1-\frac{\theta^2}{2}$, where $\theta^2 = n_x^2+n_y^2$. Substituting the above variables in \cref{eq:integralrad1} and \cref{eq:integralrad2} we get \cite{TobiasUdvikling}
\begin{multline}
f_x = \bigg\lbrace (\dot{v}_z+n_y\dot{v}_y)(n_x-v_x)\\ - 
 \left[n_y(n_y-v_y)+\left(-\frac{\theta^2}{2}-\delta v_z+\frac{1}{2\gamma^2}\right)\right]\dot{v}_x\bigg\rbrace
 g,
\end{multline}

\begin{multline}
f_y = \bigg\lbrace (\dot{v}_z+n_x\dot{v}_x)(n_y-v_y)\\ - 
 \left[n_x(n_x-v_x)+\left(-\frac{\theta^2}{2}-\delta v_z+\frac{1}{2\gamma^2}\right)\right]\dot{v}_y\bigg\rbrace
 g,
\end{multline}

\begin{multline}
fz = \bigg\lbrace \bigg[-\frac{\theta^2}{2}-\delta v_z+\frac{1}{2\gamma^2}\bigg](n_x\dot{v}_x+n_y\dot{v}_y)\\ - [n_x(n_x-v_x)+n_y(n_y-v_y)]\dot{v}_z\bigg\rbrace g,
\end{multline}
where 
\begin{equation}
g = \left(\frac{1}{2\gamma^2}+\frac{\theta^2}{2}-\delta v_z-n_xv_x-n_yv_y\right)^{-2},
\end{equation}
and the exponential phase becomes
\begin{equation}
\omega(t-\bold{n}\cdot\bold{x}) = \omega\left(\left(\frac{1}{2\gamma^2}+\frac{\theta^2}{2}\right)t-\delta z-n_x x-n_y y\right).
\end{equation}
This expansion is used in both the classical radiation integral and the following two semi-classical integrals.

In many cases the largest quantum effect to the radiation process is the photon recoil. It is a kinematical effect that can be taken into account by a simple substitution of the frequency variable in the classical photon number spectrum regardless of the details of the motion \cite{Lind91}. By substituting the frequency $\omega \rightarrow \omega^* = \omega/(1-\hbar\omega/E)$ in the classical number spectrum one gets exactly the quantum number spectrum
\begin{equation}
\frac{dN_\text{c}}{d\hbar\omega}(\omega^*) = \frac{dN_\text{q}}{d\hbar\omega}(\omega),
\end{equation}
which means that the quantum intensity spectrum becomes
\begin{equation}\label{Lindhard_substi}
\frac{dI_{\text{q}}}{d\hbar\omega}(\omega) = \frac{\omega}{\omega^*}\frac{dI_{\text{c}}}{d\hbar\omega}(\omega^*).
\end{equation}
When quantum effects related to the spin of the particles can be neglected, which is generally the case for electron/positron energies below approximately 2 TeV/Z, where Z is the charge number of the crystal, employing the substitution of the frequency variable in \cref{eq:integralrad1} according to \cref{Lindhard_substi} will reproduce the full quantum spectrum. We call this model the substitution model.

To include effects of both the quantum recoil and of the
particle spin in the radiation process we apply a result
obtained on the basis of the semi-classical method by Baier and Katkov \cite{Baier}, in which the particle motion is still
treated classically whereas the interaction with the radiation field is quantal. Belkacem, Cue, and Kimball give the semi-classical spectrum for a general trajectory as (first seen in \cite{belkacem_1985} and derivations shown in \cite{TobiasUdvikling} \cite{kimball_1986}):
\begin{equation}\label{cue}
\frac{dI}{d\omega d\Omega}=\frac{e^2}{4\pi^2}\left(\frac{E^{*2}+E^2}{2E^2}\left| I \right|^2 + \frac{\omega^2}{2E^2\gamma^2}\left| J \right|^2\right),
\end{equation}
where $E^* = E-\hbar\omega$ and $I$ and $J$ are given by
\begin{equation}\label{Iterm}
I = \int_{-\infty}^{\infty} \frac{\bm{n}\times[(\bm{n}-\bm{\beta})\times\dot{\bm{\beta}}]}{(1-\bm{n}\cdot\bm{\beta})^2}e^{i\omega^*(t-\bm{n}\cdot\bm{r})}dt,
\end{equation}
\begin{equation}
J = \int_{-\infty}^{\infty} \frac{\bm{n}\cdot\dot{\bm{\beta}}}{(1-\bm{n}\cdot\bm{\beta})^2}e^{i\omega^*(t-\bm{n}\cdot\bm{r})}dt.
\end{equation}
We call this the BCK model.

The substitution model and the BCK model are both derived in a first order approximation in the interaction with the radiation field, which means that they apply for single photon emission and only for pure Lorentz force trajectories. Due to the energy dependent phase factor, the radiation spectrum from a particle following a long trajectory, in which it loses energy, will be incorrect. If the particle loses energy along the trajectory, the substituted frequency variable becomes time dependent and the result depends on the initial phase, which is non-physical. 

To account for this we divide the trajectory into smaller sections in which the energy loss of the particle has no effect on the spectrum. We then fix the energy of the particle in the radiation integral to the initial value it has when entering the given section. The initial conditions of the particle when entering a new section is then the final conditions from the previous section. The full spectrum for an entire trajectory is the sum of the spectra from each section.
The section length is chosen by checking that the spectrum from a pure Lorentz force trajectory is the same as a trajectory where the particle loses energy due to radiation emission. For 50 GeV positrons channeled in the (110) plane of Si this length will be 0.1 mm. One also has to ensure that the section length is longer than the formation length \cite{
andersen2014,formation} $l_f =2\gamma^2(E-\omega)/E\omega$, which in the case of a 50 GeV positron emitting a 5 GeV photon is $l_f = 71$ nm.

\section{Crystal channeling}
Channeling of charged particles in crystals has been studied in great detail \cite{Baier}, and has been shown to significantly affect radiation emission from charged particles \cite{Ulrikcrystals}. Here we describe the physics of the channeling process which is to be implemented in the C++ code. 

The particles move in the electric field of the crystal lattice, described in our model by the thermally averaged Doyle-Turner continuum potential \cite{Doyle,JUAndersen}. The potential pertaining to a single plane and a single axial atomic string, respectively, for a particle with charge $e$ is given by \cite{SørenPapeBending}
\begin{small}
\begin{equation}\label{eq:plane}
U(x) = 2\sqrt{\pi}e^2a_0Nd_p\sum_{i=1}^4\frac{a_i}{\sqrt{B_i+\rho^2}}\text{exp}\left(\frac{-x^2}{B_i+\rho^2}\right),
\end{equation}
\begin{equation}\label{eq:string}
U(r) = \frac{2e^2a_0}{d}\sum_{i=1}^4\frac{a_i}{B_i+\rho^2}\text{exp}\left(\frac{-r^2}{B_i+\rho^2}\right).
\end{equation}
\end{small}Here $Z_1$ is the charge number of the projectile, $\rho$ is the two dimensional thermal vibration amplitude at room temperature, $a_0$  is the Bohr radius, $d_p$ Å is the distance between two planes, $d$ is the mean distance between atoms in a string, $x$ and $r$ is the distance to a plane or a string respectively and $B_i = b_i/4\pi^2$. The values of $b_i$ and $a_i$ are fitting parameters determined by electron atomic scattering factors. These have first been measured by Doyle and Turner \cite{Doyle} for the potentials in \cref{eq:plane} and \cref{eq:string} and later for more elements by Peng \cite{Peng}.

Multiple Coulomb scattering on individual nuclei and electrons in the crystal are accounted for by Monte Carlo based successive elastic changes in direction of motion in each integration step. 
The mean squared scattering angle per length traveled is the sum of the mean squared scattering angles on electrons and nuclei \cite{ElectronScat}
\begin{equation}
\frac{\overline{\Delta\theta^2}}{\Delta z} = \frac{\overline{\Delta\theta^2}_n}{\Delta z}+\frac{\overline{\Delta\theta^2}_e}{\Delta z}.
\end{equation}
Here the mean squared scattering angle due to collision with nuclei is given by \cite{PDG}
\begin{equation}
\frac{\overline{\Delta\theta^2}_n}{\Delta z} = \frac{\epsilon_s^2}{E^2 }\frac{1}{X_0} n(r),
\end{equation}
where $X_0$ is the radiation length, $E$ is the energy of the particle, $\epsilon_s = 10.6$ MeV and 
\begin{equation}
n_{p}(r) = N\frac{d_p}{\sqrt{\pi \rho^2}}\text{exp}\left(-\frac{r^2}{\rho^2}\right),
\end{equation}  
\begin{equation}
n_{s}(r) = N\frac{1}{{\pi \rho^2Nd}}\text{exp}\left(-\frac{r^2}{\rho^2}\right),
\end{equation}
is the local density of nuclei pertaining to a single plane or string respectively \cite{Scattering}. One sums the density from all nearby planes or strings to get the total local density. Here $N$ is the mean nuclei density in the material and $r$ is the distance to the plane or the string.
The mean squared scattering angle due to collision with valence electrons is given by \cite{ElectronScat}
\begin{equation}\label{eq:elscat}
\frac{\overline{\Delta\theta^2}_{el}}{\Delta z} = \frac{\pi r_0^2}{\gamma^2\beta^4}\left[\log\left(\frac{2m_e\gamma^2\beta^2}{I}\right)-\beta^2\right]n_{el}(r),
\end{equation}
where $r_0$ is the classical electron radius, $m_e$ is the electron rest mass, $I$ is the average ionization energy of the target atoms and
\begin{multline}
n_{el,p}(r) = \frac{d_p}{\pi}\sum_{i=1}^4\frac{a_i^{(X)}}{\sqrt{B_i^{(X)}+\rho^2}}\text{exp}\left(\frac{-r^2}{B_i^{(X)}+\rho^2}\right)\\+c^{X}n_p(r),
\end{multline}

\begin{multline}
n_{el,s}(r) = \frac{1}{\pi d}\sum_{i=1}^4\frac{a_i^{(X)}}{B_i^{(X)}+\rho^2}\text{exp}\left(\frac{-r^2}{B_i^{(X)}+\rho^2}\right)\\
+c^{X}n_s(r),
\end{multline}
is the local electron density \cite{Elektrondensity} for a plane and a string respectively. Here the values of $a_i^{(X)}$, $b_i^{(X)}$ and $c^{(X)}$ are the X-ray scattering factors also measured by Doyle and Turner \cite{Doyle}.
In \cref{eq:elscat} we have removed a factor of 2 to ensure that the radiation yield from scattering on nuclei and electrons differs by a factor of $1/Z$ \cite{PDG}. 

To account for the loss of energy due to radiation emission we include the Landau-Lifshitz (LL) force in the equation of motion of the particles, which gives \cite{Landau}
\begin{equation}
\label{iLL_eq}
\begin{split}
m\frac{d u^{\mu}}{ds}=&eF^{\mu\nu}u_{\nu}+\frac{2}{3}e^2\left[\frac{e}{m}(\partial_{\alpha}F^{\mu\nu})u^{\alpha}u_{\nu}\right.\\
&\left.+\frac{e^2}{m^2}F^{\mu\nu}F_{\nu\alpha}u^{\alpha}+\frac{e^2}{m^2}(F^{\alpha\nu}u_{\nu})(F_{\alpha\lambda}u^{\lambda})u^{\mu}\right].
\end{split}
\end{equation}
Here, $m$ and $e$ denote the particle mass and charge respectively, $F^{\mu\nu}$ is the external electromagnetic field tensor, $u^{\mu}$ is the particle four-velocity, and $s$ its proper time, in units where $c = 1$.
Because the LL equation depends on the electric field, it does not react to the instantaneous "kick" the particle get when it undergoes multiple Coulomb scattering, as it is put in by hand. For a uniform constant density of nuclei and electrons the radiation spectrum evaluated should reproduce the Bethe-Heitler bremstrahlung spectrum, which means that  the energy loss due to bremstrahlung is not accounted for. We implement this by the scheme developed in \cite{TobiasSpredning}, where the energy loss of the particle due to one deflection is given by
\begin{equation}
dE_{MS} = \frac{m_{el}\alpha}{2\pi}\gamma^3\theta^2.
\end{equation}
Here $\theta$ is the Monte-Carlo sampled deflection angle in every step from multiple Coulomb scattering.

\section{Code implementation}
The entire program can be divided into two parts; calculating the trajectory and calculating the radiation integral, given the trajectory. The two parts work independently and the radiation integrals are evaluated on the GPU while the trajectories are evaluated on the CPU. The reason why the trajectories are not evaluated on the GPU is because the task of solving the equation of motion is a serial task which only has to be done once for each particle and the number of particles usually does not exceed 2000 per simulation. This means that we are not even close to saturating high end GPUs.

 \begin{figure}
 \includegraphics[width=\linewidth]{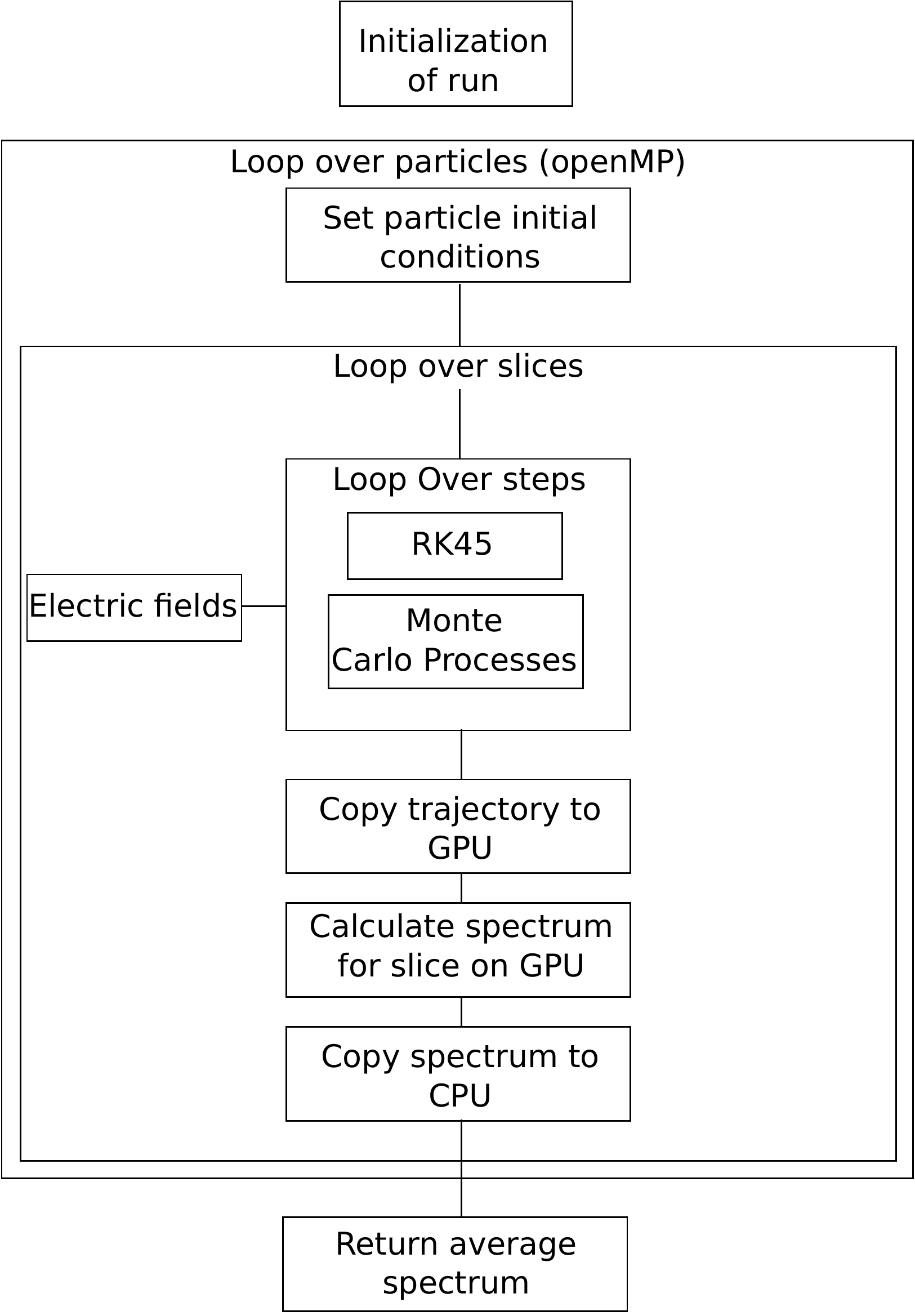}
 \caption{Flow diagram of the code. As the crystal is divided into sections in which the spectrum is evaluated, we loop over each of these sections, which we call a slice. A step is going from one point in the trajectory to the next.}
 \label{fig:codeflow}
 \end{figure}

On \cref{fig:codeflow} a diagram of the code structure is shown. This outlines the basic steps that are taken in the process of evaluating a radiation spectrum.
The trajectories are evaluated with a Runge-Kutta (RK45) Ordinary Differential Equation (ODE) solver which features step size modulation. We have implemented this method including a lower bound on the modulated step size in order to avoid step sizes converging to zero due to stiffness of the differential equation. The stiffness of the differential equation only affects the part of the trajectory for electrons very close to the string in the axial case. The error imposed by setting a fixed lower step size, which is reached in only a few steps during a simulation of 1000 particles, we estimate is much lower than the inaccuracy of the theoretical models used to calculate the fields and multiple scattering.
Since Monte Carlo based processes are included, like multiple scattering, we first do one step with the ODE solver, and afterwards change the trajectory by hand, based on the Monte Carlo processes. This step can be modulated in size within the solver but it returns only the trajectory at the point in time specified by the step. In this way we can control the resolution of the trajectory, this is vital when calculating the radiation spectrum, as for high-frequency photon emission the radiation integral oscillates much more rapidly and more points in time are required for the integral to converge.

We have written a library that evaluates the effective electric field and gradient of the field as a function of position within the crystal and which crystal type/orientation we are using. This library features the placement of strings and planes for FCC and BCC crystals in the (111), (110), (100), $\langle111\rangle$, $\langle110\rangle$ and $\langle100\rangle$ orientations. The elements which have been implemented are Silicon, Diamond, Germanium and Tungsten, with easy implementation of new elements which have the same crystal structure.

When the trajectory is evaluated, it is copied onto the GPU memory on which the spectrum pertaining to the trajectory is evaluated and copied back to the CPU. By only copying the trajectory and the spectrum once we limit the amount of data copied between GPU and CPU and therefore reach more than 99\% workload on the GPU during the entire simulation. 

To produce a spectrum of the type $dI/d\omega$ we need to evaluate the spatial integral over $d \Omega$ for every frequency $\omega$ in the spectrum. We therefore have to evaluate the integral over the same trajectory $N_x \times N_y \times N_\omega$ times, where $N_x $ and $N_y $ is the number of points in the spatial integral in the $x$ and $y$ direction respectively and $N_\omega$ is the number of points in the spectrum. As an example, it is not unreasonable to have $N_x = N_y = 50$ and $N_\omega = 100$, when multiple scattering is included, which amounts to $0.25\cdot10^6$ integrals over the same trajectory. When multiple scattering is included the spatial features of the spectrum are spread out and only 50 points in both spatial dimensions are needed, but if multiple scattering is not included, and the particle follows a periodic motion, the spatial features become very narrow, and depending on how many periods the integral contains, we could need more than 3000 points in both spatial dimension. This would amount to $900\cdot10^6$ individual integrals over the same trajectory, which at this point begin to approach the limit of how much memory is accessible on the GPU. The average spectrum of a beam of particles would then be the average spectrum of many particles with different initial conditions entering the crystal.

The above point is also one of the reasons why we evaluate the trajectories on the CPU individually, and only evaluate the spectrum pertaining to one trajectory on the GPU at a time. The GPU is saturated with work by evaluating a spectrum from just one particle, which means that we do not lose much performance by evaluating trajectories on the CPU. To always keep the GPU saturated by work we use openMP to evaluate trajectories in parallel on the CPU, each CPU thread then instructs the GPU to evaluate the spectrum pertaining to its trajectory. This means that the GPU can be working on the spectrum from one CPU thread, while another CPU thread is evaluating another trajectory. Depending on which energy, crystal orientation and number of integrals per trajectory we need, the time it takes to evaluate a trajectory can either be faster or slower than evaluating a spectrum. So it can in some cases affect the performance a lot if you are balancing between which process is the fastest. Evaluating a trajectory for a planar oriented crystal is generally faster by up to an order of magnitude than evaluating a trajectory in the axial case.  

The workload on the GPU is distributed such that each thread on the GPU evaluates one integral and saves the result in a cube (3 dimensional matrix corresponding to the dimensions  $N_x$, $ N_y $ and $ N_\omega$). This cube is copied back onto the CPU where we do the final sum over the spatial integrals evaluated by the GPU and produce the final spectrum. 

As already mentioned the radiation spectra rely on small differences between large numbers, this means that we need high precision on the data type used, and we are required to use double precision numbers when calculating the trajectory and the radiation spectrum. This is required even though we do a series expansion as explained in \cref{radiationintegral}.

A version of the code can be found in the repository on \footnote{https://gitlab.au.dk/au483748/channelingradiation.git}.

\subsection*{Test of integral implementation}
 \begin{figure*}[t]
 \includegraphics[width=\linewidth]{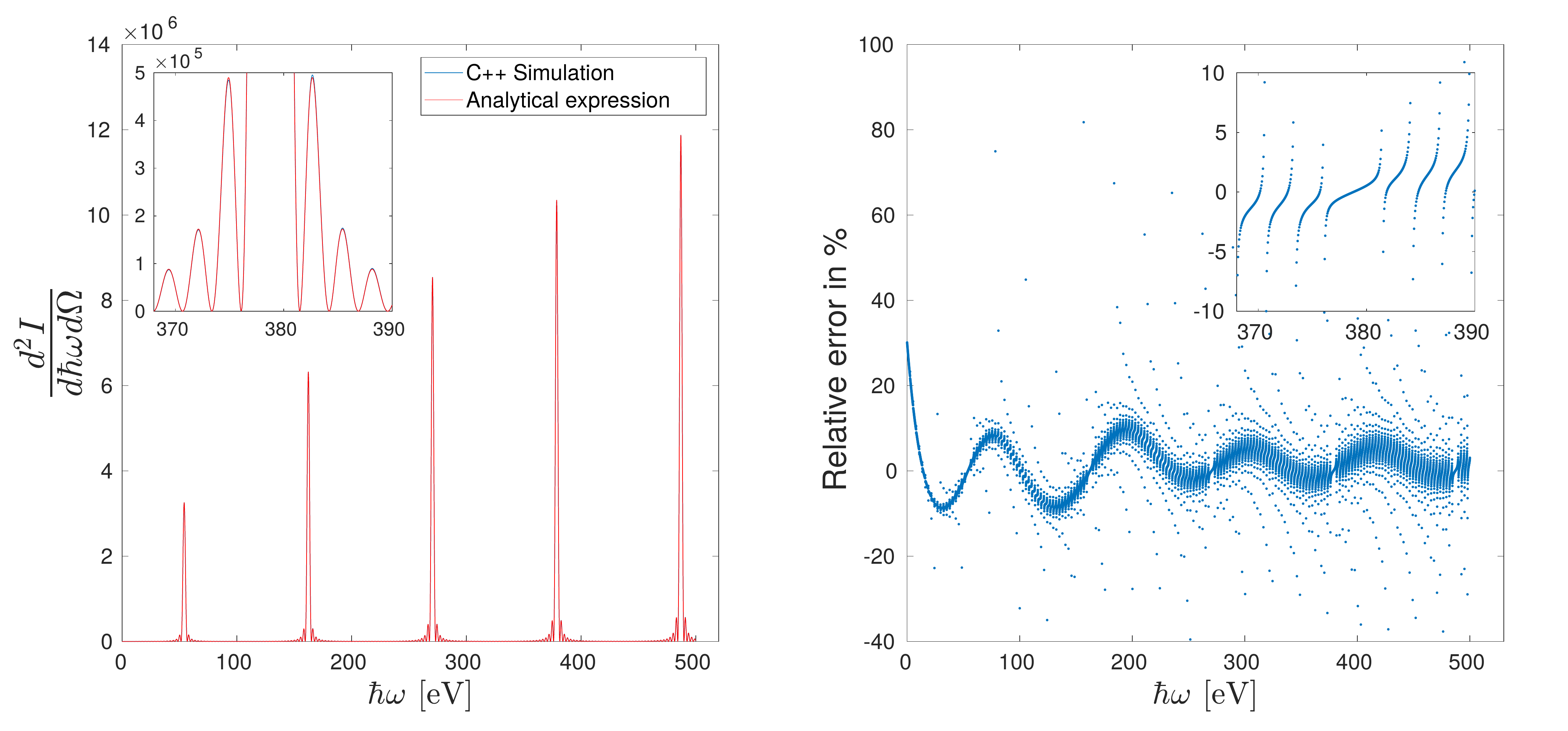}
 \caption{Radiation Spectrum from a 2 GeV electron in an undulator in the forward traveling direction, with an undulator period length $\lambda = 53$ mm, $N = 20$ periods and a magnetic field strength $B_0 = 1$ T. The figure on the left shows the radiation spectrum evaluated using the CUDA program (blue), together with an analytical solution to the same problem (red), found in\cite{undulator}. The inserted figure is a zoom on the peak at 380 eV and even in this case the blue curve is hardly visible due to the almost perfect match with the red curve. The figure on the right shows the relative error between the two solutions. The inserted figure is a zoom of what corresponds to the peak around 380 eV.}
 \label{fig:undulattest}
 \end{figure*}
 \begin{figure*}[]
 \includegraphics[width=\linewidth]{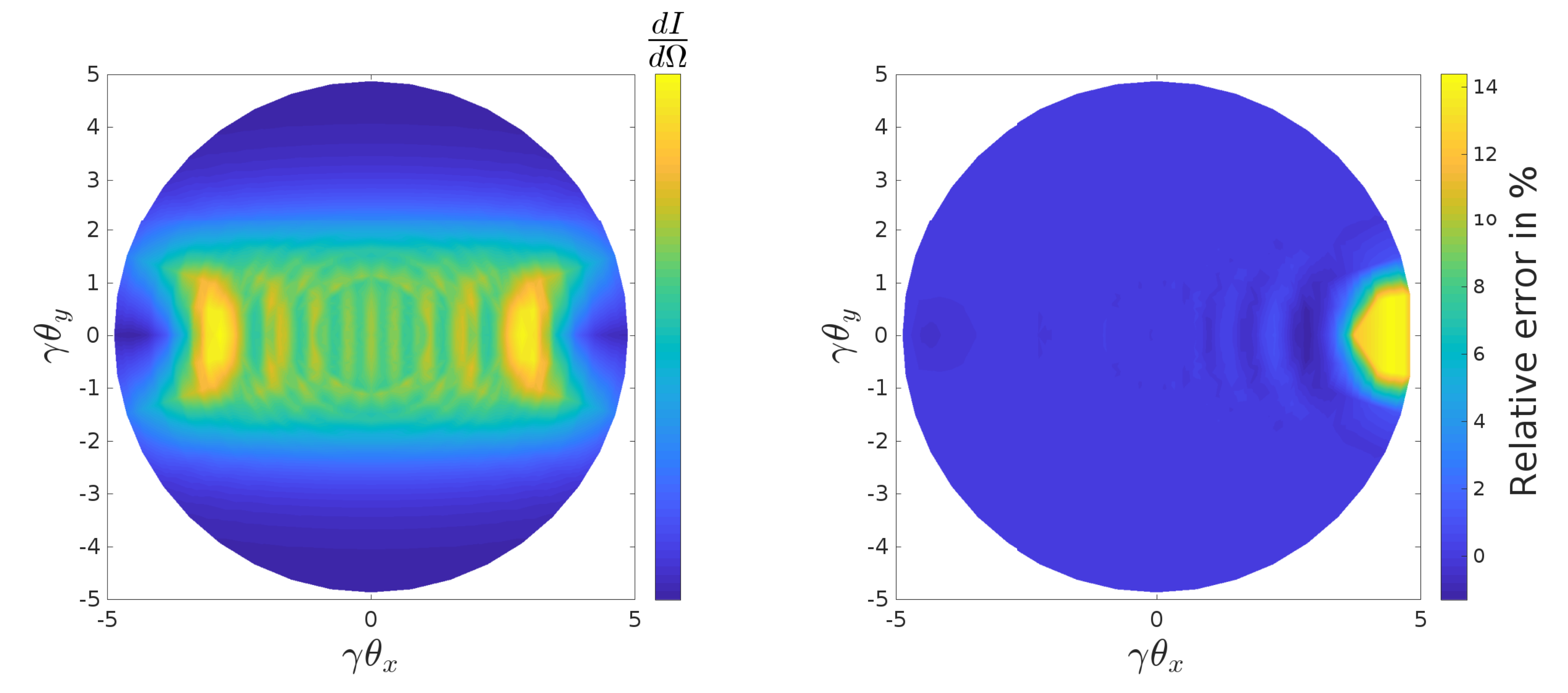}
 \caption{$\frac{dI}{d\Omega}$ from a 2 GeV electron in an undulator with an undulator period length $\lambda = 53$ mm, $N = 20$ periods and a magnetic field strength $B_0 = 1$ T. The integrated photon energy range is 0-500 eV. Left figure is the solution found by the CUDA program. Right figure is the relative error in percent between the solution found by the CUDA program and the analytical solution found in\cite{undulator}.}
 \label{fig:undulattest2}
 \end{figure*}

To test the numerical implementation of \cref{eq:integralrad1}, we evaluated the radiation emitted from an electron traversing a plane strong undulator in the $z$-direction, with a magnetic field in the y direction:
\begin{align}
H_y = B_0\text{sin}\left(\frac{2\pi z(t)}{\lambda}\right).
\end{align}  
In the test we used a magnetic field strength $B_0 = 1$ T, an undulator period length $\lambda = 53$ mm, an electron at 2 GeV and $N=20$ periods. The resulting radiation is compared to the analytical expression found in \cite{undulator}. On \cref{fig:undulattest} the radiation evaluated with the CUDA program in the forward direction is shown together with an analytical solution to the problem and their relative error. We clearly see good agreement between the two solutions. In vicinity of the peaks the error drops to around 1-2\% and in the areas where there is no peak, we see rather large error oscillations up to 20\%, and even a few larger than 100\% which are not shown in the figure. These large errors occur in areas where the functions are very small. While the analytical solution might be zero by construct in the valleys, small numerical errors in the program result in large deviations from the analytical solution. These error patterns are therefore expected.

We also evaluated the energy radiated per solid angle $dI/d\Omega$ in an area with a maximal opening angle of $5/\gamma$, numerically integrating \cref{eq:integralrad1} with respect to $\hbar\omega$ from 0-500 eV. This is shown on \cref{fig:undulattest2}, together with the relative error in percent between the solution found by the CUDA program and the analytical solution. We again see good agreement between the CUDA program and the analytical solution, with relative errors within a few percent, except in a small area. In \cref{fig:undulattest2} we only show relative errors up to 14\% in order to highlight the smaller errors, but in the center of the area with the large errors lies a few points with up to 200\% error which is not shown. These large errors, we believe, are again results of small numerical errors in the CUDA program. The function values we compare in the area are very small, and a small numerical error will result in large differences between the functions.

From this test we conclude that the CUDA program written to evaluate radiation emitted from accelerating charged particles given the particle trajectory works, and could be used to evaluate the radiation emitted from arbitrary motion, e.g. positrons and electrons channeled in crystals. We therefore proceed to testing this as described in the next section.

\subsection*{Test of trajectory implementation}

\begin{figure}[]
 \includegraphics[width=\linewidth]{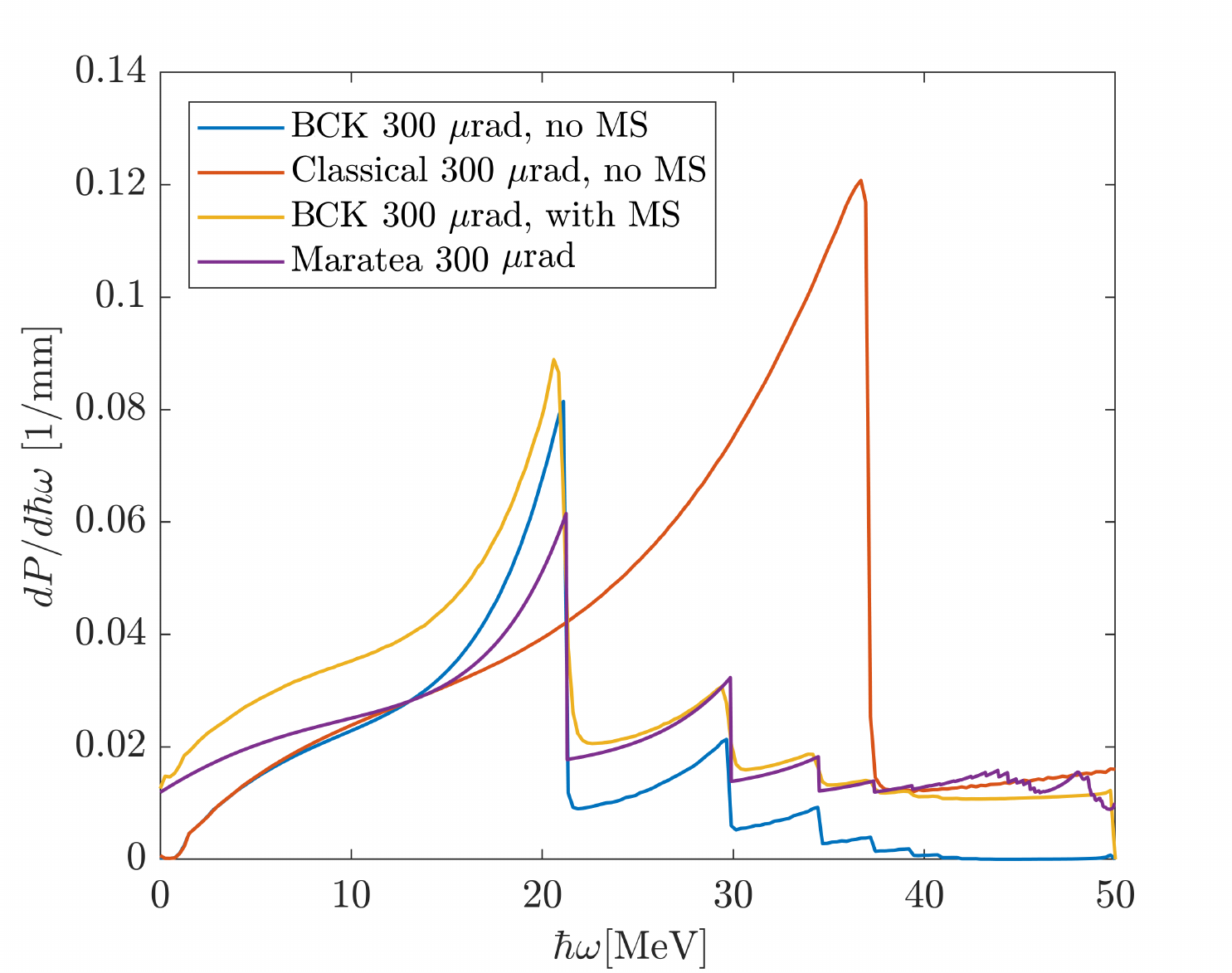}
 \caption{Radiation spectrum from a 50 GeV positron traversing the (110) plane of Si at an angle of 300 $\mu$rad. The label MS tells whether multiple Coulomb scattering is included or not. The classical and BCK model are both the result from the CUDA/C++ program, and the curve labeled Maratea is the result of a quantum mechanical calculation done by Allan H. S\o{}rensen at Aarhus University, using the model described in \cite{coherentbremstrahlung,coherentbremstrahlung2}.}
 \label{fig:maratea}
 \end{figure}

We test the implementation of the continuum potential by evaluating the radiation emitted from a 50 GeV positron traversing a (110) oriented Si crystal, initially moving at an angle of 300 $\mu$rad with respect to the crystal planes with an initial position between two planes. These simulations are compared to quantum mechanical calculations using the formulation of coherent bremstrahlung found in \cite{coherentbremstrahlung,coherentbremstrahlung2}. Here the particle is treated as a plane wave traversing the crystal planes with an angle. This model breaks down for small angles, while the continuum model breaks down for large angles. Even though the models break down in different regimes, the position of the harmonic peaks is correct, as this is a kinematic effect related to the frequency of which the particles traverse a plane. On \cref{fig:maratea} the spectra using the classical radiation integral and the BCK model is shown together with the coherent bremstrahlung calculation which we call Maratea, produced by Allan H. S\o{}rensen from Aarhus University. It is clear that the classical model no longer agrees with the quantum calculations, since effects like the photon recoil is neglected. The discrepancy between the classical model and the BCK model starts to show even at 5 GeV photon energies (10\% of the incoming particle energy). We see good agreement of the location of the harmonic peaks between the BCK and the Maratea model. It is clear that including multiple Coulomb scattering is necessary to describe the level difference at high photon energy, which relates to incoherent bremstrahlung.  

\begin{figure}[]
 \includegraphics[width=\linewidth]{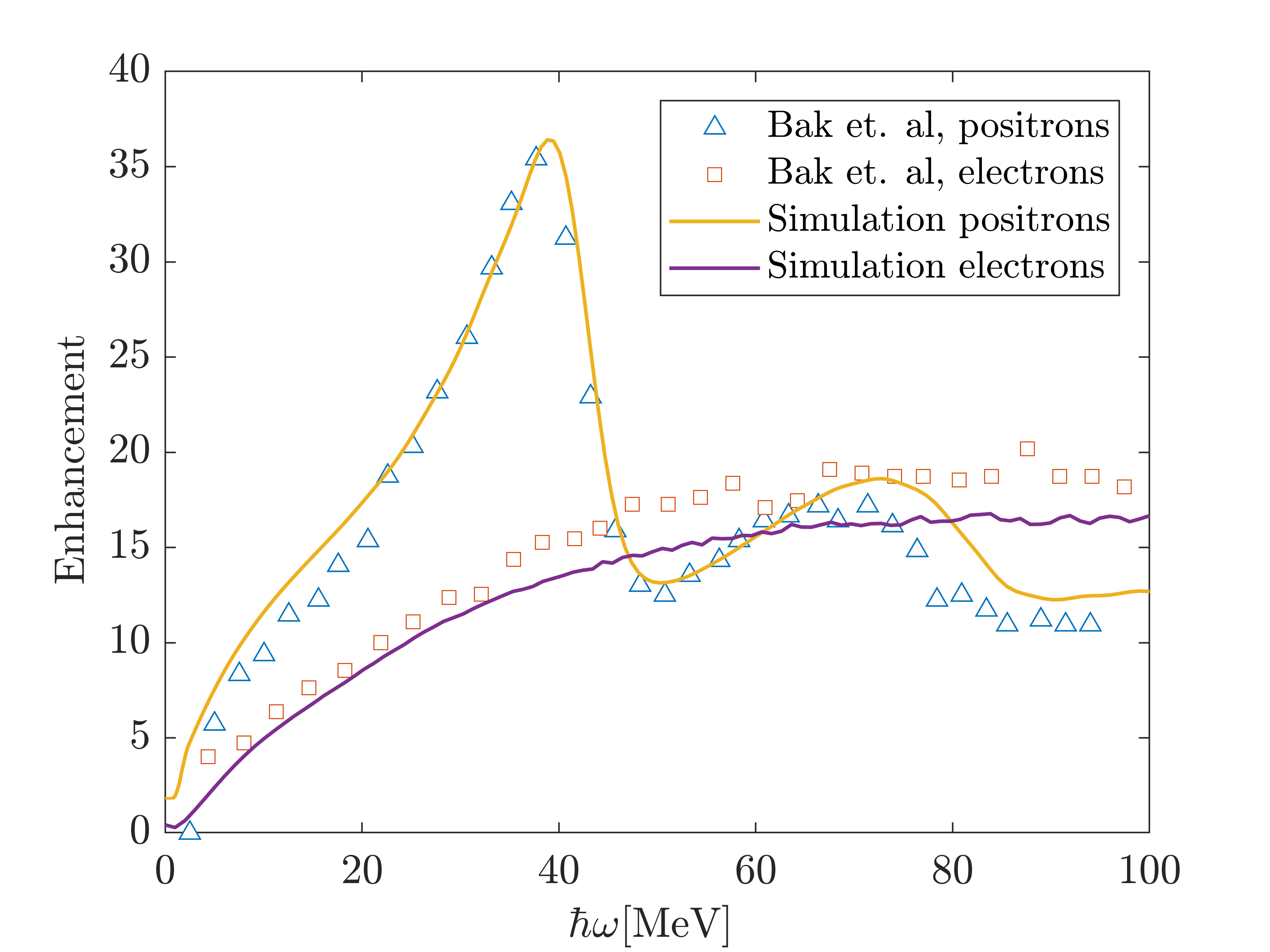}
 \caption{Radiation spectra from 6.7 GeV positrons and electrons traversing the (110) plane of a 0.105 mm thick Si crystal. Entry angles of the particles in the simulation are uniformly distributed between $\pm110$ $\mu$rad. The squares and triangles are experimental data points digitized from \cite{Baketal}.}
 \label{fig:baketal}
 \end{figure}

A final test of the implementation of radiation emission when multiple Coulomb scattering in oriented crystals is included, involve evaluating the radiation spectrum from 6.7 GeV positrons and electrons traversing the (110) plane of a 0.105 mm thick Si crystal. On \cref{fig:baketal} we show a comparison of the theoretical spectrum with experimental data measured by the authors from \cite{Baketal}. By considering the beam divergence and angular resolution in the experimental cuts, the entry angle of the simulated particles is chosen to follow a uniform distribution of angles between $\pm110$ $\mu$rad. We see good agreement between simulation and experiment, even when looking at the spectral features of the radiation from the positrons.

\section{GPU vs CPU}
In this section we compare computation times of the radiation integrals between different GPUs and CPUs. We will be using a variety of GPUs to also illustrate the performance increase between GPUs.
The two CPUs used in the test are the Intel 7820x processor, which features 16 parallel cores and the second CPU is the Xeon Gold 6140 processor, featuring 36 parallel cores. The GPUs used in the test are the GeForce 1080ti, Titan V and the Quadro GP100. The code used for the CPU's are nearly identical to the GPU version, as the data structure and function evaluations are the same. Since the data structure used are only c-style arrays and the fact that we only use low-level math functions like sines and square roots, the GPU code is also optimized well for CPU's. The only difference is that no data is transferred between GPU and CPU, and that now only one thread is working on the spectrum from a single trajectory if many particles are simulated. In the following tests we only simulate one particle, which means that all threads available on the CPU will be working on the spectrum just like in the GPU case, evaluating one integral each.

 \begin{figure}[]
 \includegraphics[width=\linewidth]{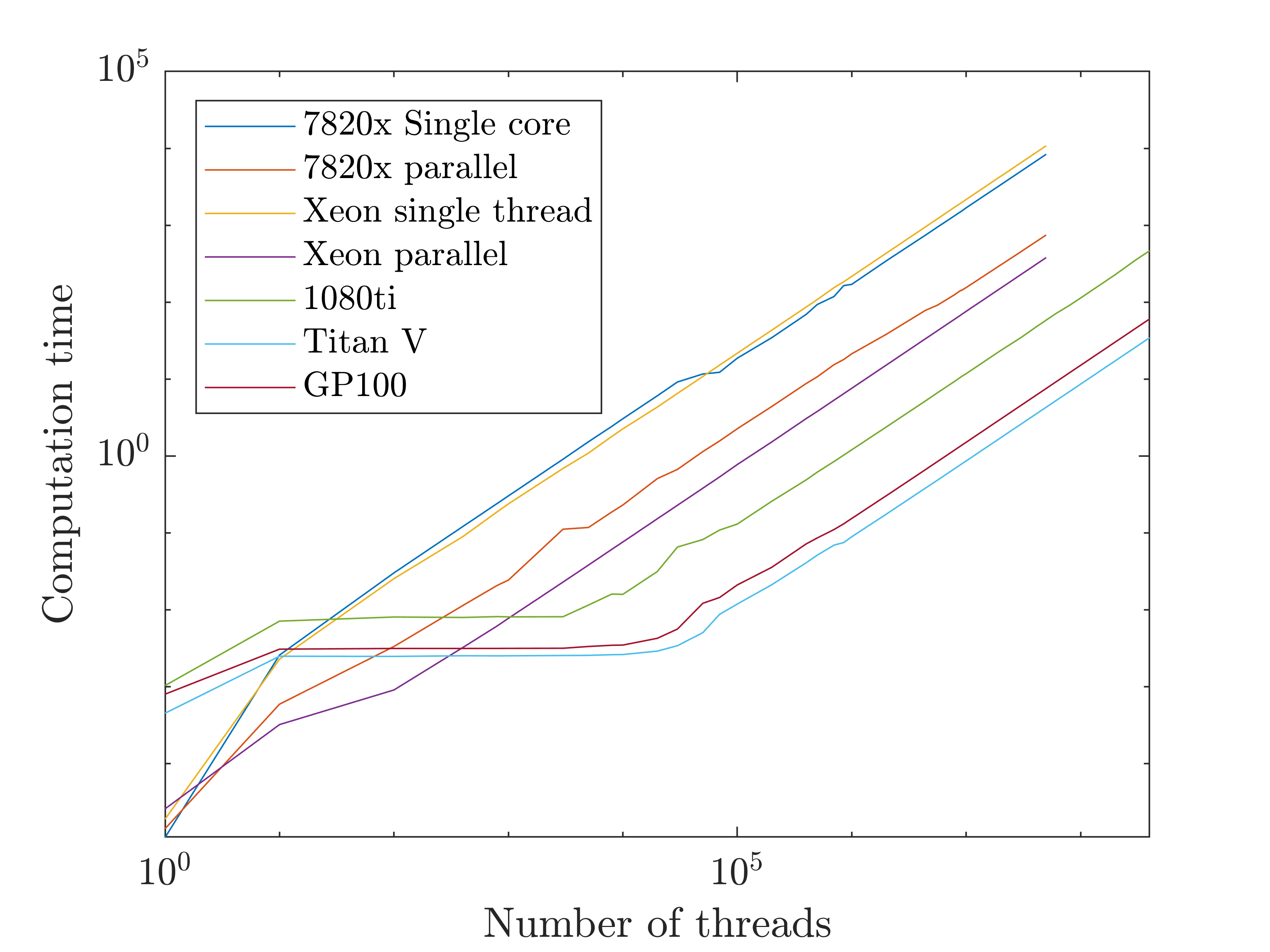}
 \caption{Computation time as a function of the number of evaluated radiation integrals (\cref{eq:integralrad1}), each integral has 3000 points in time. The 7820x CPU has 16 parallel cores, and the Xeon processor has 36 parallel cores.}
 \label{fig:comptime}
 \end{figure}
The test involves measuring the computation time of a certain amount of radiation integrals of the form in \cref{eq:integralrad1}. Each integral consist of 3000 points in time, which corresponds to $\approx0.01$ mm thick planar oriented crystal. To do the spatial integration of $d\Omega$, one has to calculate the integral \cref{eq:integralrad1} for a certain number of points in space, which in the case of the spectra in \cref{fig:maratea} amounts to $5\cdot 10^6$ points in space, since the spatial spectral features become very narrow when propagating a particle along the same oscillatory motion for an extended time. Then to produce a spectrum one has to also calculate each integral for each photon frequency in the spectrum, corresponding to $5\cdot 10^8$ integrals in total. It is therefore necessary to see how each chip performs as a function of the number of integrals.

On \cref{fig:comptime} the computation time for each chip is shown. The processors are run in single core mode, doing every integral sequential, and in parallel mode, meaning that their 16 and 36 cores respectively are working in parallel. As expected the CPUs have faster computation time for very low number of threads, we also see that the single core processors are faster than the parallel processor, which is due to the overhead of starting a parallel process. 
We see a flat trend for each GPU for $\approx 10^4$ threads, this is due to the amount of cores available on the GPUs not saturated with working threads. The number of threads needed before each GPU is saturated is different for each GPU and we see that the first GPU to reach saturation in threads is the 1080ti which is expected. What is perhaps not expected is that the most expensive GPU, the Quadro GP100, reaches saturation before the Titan V. After saturation the computation time for the GPUs become linear in number of threads, as the CPUs. From \cref{fig:comptime} we also see that the number of threads needed for the Xeon CPU and the Titan V GPU to reach identical performance is only 350, after which the Titan V is faster.

 \begin{figure}[]
 \includegraphics[width=\linewidth]{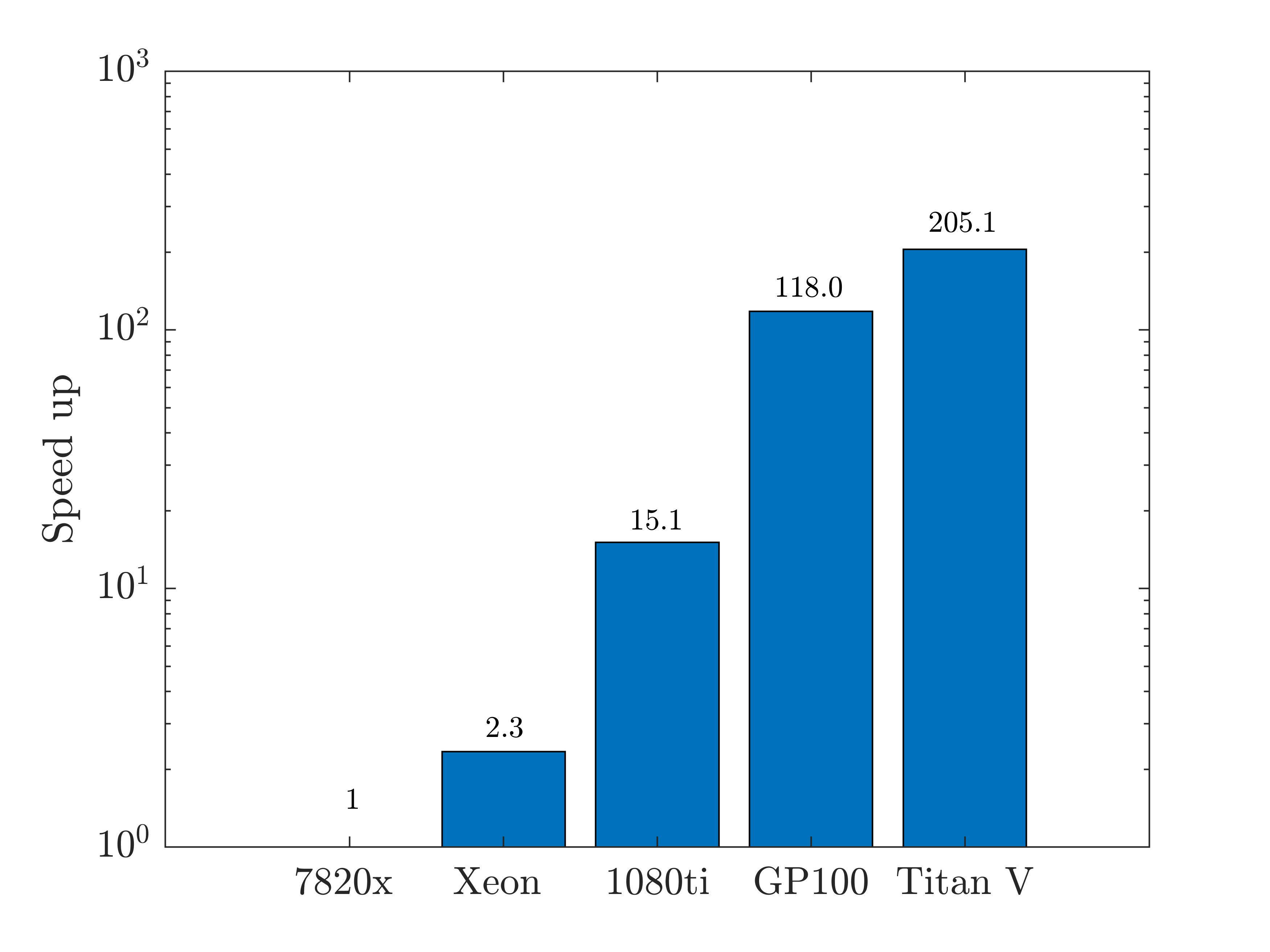}
 \caption{Speed up with respect to the 7820x processor running in parallel mode. The speed up value for each chip is listed above their respective bar.}
 \label{fig:speedup}
 \end{figure}

On \cref{fig:speedup} the speed-up of each chip is shown compared to the 7820x processor in parallel mode. Here each chip is fully saturated by threads. As expected we see a factor of $\approx 2$ between the 36 cores from the Xeon processor and the 16 cores on the 7820x processor, while we reach a speed up of 205.1 for the Titan V. 

This shows that if one had a parallel job capable of running effectively on GPUs, a single Titan V could replace a computer cluster running almost 100 nodes, each equipped with a 36 core Xeon processor. 

One thing also worth noting is that these integrals require double precision operations. The architecture on the Titan V and the GP100 allows for much faster double precision calculations than earlier models like the 1080ti \cite{1080ti} and even the new Titan RTX which is not built to support effective double precision calculations \cite{turing}. Choosing a GPU for a specific job therefore depends on the datatype used in the calculation, which becomes more important when calculations require double precision.

\section{Concluding remarks and future implementations}
The current version of the CUDA code is a program capable of evaluating both classical and quantum mechanical radiation spectra from charged relativistic spin 0 and 1/2 particles following arbitrary classical trajectories. These calculations are done on a single GPU and an even higher speedup could be achieved using a cluster of GPUs. The bottleneck in computation time is calculating the actual trajectory of particles when simulating particles traversing oriented single crystals. The C++ program feature straight axial and planar oriented crystal trajectory simulations for Tungsten, Diamond, Germanium and Silicon. Every particle trajectory is calculated in parallel on the CPU and an implementation where many particle trajectories are evaluated in parallel on the GPU is of interest. Also of interest is simulation of bent crystals, which can be used as crystalline undulators or as a beam extraction device for accelerators like the LHC.

It is clear from our analysis that GPUs can be a very powerful tool when extensive calculations become impossible due to computation time. The evolution of computation performance of GPUs the past decade has pushed the boundaries of what is numerically possible and we believe that GPUs are the future of computation in many areas of science, not only physics.  

The numerical results presented in this work were partly obtained at the Centre for Scientific Computing, Aarhus http://phys.au.dk/forskning/cscaa and by support from Nvidias GPU grant program.


\begin{thebibliography}{30}%
\makeatletter
\providecommand \@ifxundefined [1]{%
 \@ifx{#1\undefined}
}%
\providecommand \@ifnum [1]{%
 \ifnum #1\expandafter \@firstoftwo
 \else \expandafter \@secondoftwo
 \fi
}%
\providecommand \@ifx [1]{%
 \ifx #1\expandafter \@firstoftwo
 \else \expandafter \@secondoftwo
 \fi
}%
\providecommand \natexlab [1]{#1}%
\providecommand \enquote  [1]{``#1''}%
\providecommand \bibnamefont  [1]{#1}%
\providecommand \bibfnamefont [1]{#1}%
\providecommand \citenamefont [1]{#1}%
\providecommand \href@noop [0]{\@secondoftwo}%
\providecommand \href [0]{\begingroup \@sanitize@url \@href}%
\providecommand \@href[1]{\@@startlink{#1}\@@href}%
\providecommand \@@href[1]{\endgroup#1\@@endlink}%
\providecommand \@sanitize@url [0]{\catcode `\\12\catcode `\$12\catcode
  `\&12\catcode `\#12\catcode `\^12\catcode `\_12\catcode `\%12\relax}%
\providecommand \@@startlink[1]{}%
\providecommand \@@endlink[0]{}%
\providecommand \url  [0]{\begingroup\@sanitize@url \@url }%
\providecommand \@url [1]{\endgroup\@href {#1}{\urlprefix }}%
\providecommand \urlprefix  [0]{URL }%
\providecommand \Eprint [0]{\href }%
\providecommand \doibase [0]{http://dx.doi.org/}%
\providecommand \selectlanguage [0]{\@gobble}%
\providecommand \bibinfo  [0]{\@secondoftwo}%
\providecommand \bibfield  [0]{\@secondoftwo}%
\providecommand \translation [1]{[#1]}%
\providecommand \BibitemOpen [0]{}%
\providecommand \bibitemStop [0]{}%
\providecommand \bibitemNoStop [0]{.\EOS\space}%
\providecommand \EOS [0]{\spacefactor3000\relax}%
\providecommand \BibitemShut  [1]{\csname bibitem#1\endcsname}%
\let\auto@bib@innerbib\@empty
\bibitem [{\citenamefont {Jackson}(1998)}]{jacksons}%
  \BibitemOpen
  \bibfield  {author} {\bibinfo {author} {\bibfnamefont {J.~D.}\ \bibnamefont
  {Jackson}},\ }\href@noop {} {\emph {\bibinfo {title} {Classical
  Electrodynamics}}},\ \bibinfo {edition} {3rd}\ ed.\ (\bibinfo  {publisher}
  {John Wiley and Sons, Inc},\ \bibinfo {year} {1998})\BibitemShut {NoStop}%
\bibitem [{gp1(2016)}]{gp100}%
  \BibitemOpen
  \href@noop {} {\emph {\bibinfo {title} {NVIDIA TESLA P100 GPU
  ACCELERATOR}}},\ \bibinfo {organization} {Nvidia} (\bibinfo {year} {2016}),\
  \bibinfo {note} {data sheet}\BibitemShut {NoStop}%
\bibitem [{gv1(2018)}]{gv100}%
  \BibitemOpen
  \href@noop {} {\emph {\bibinfo {title} {NVIDIA TESLA V100 GPU
  ACCELERATOR}}},\ \bibinfo {organization} {Nvidia} (\bibinfo {year} {2018}),\
  \bibinfo {note} {data sheet}\BibitemShut {NoStop}%
\bibitem [{tur()}]{turing}%
  \BibitemOpen
  \href@noop {} {\emph {\bibinfo {title} {NVIDIA TURING GPU ARCHITECTURE -
  Graphics Reinvented}}},\ \bibinfo {organization} {Nvidia},\ \bibinfo {note}
  {wP-09183-001\_v01}\BibitemShut {NoStop}%
\bibitem [{Note1()}]{Note1}%
  \BibitemOpen
  \bibinfo {note} {CUDA is a C/C++ parallel computing platform designed to
  implement the use of GPUs for general purpose programming. This allows the
  user to launch thousands of parallel processes.
  https://www.geforce.com/hardware/technology/cuda}\BibitemShut {NoStop}%
\bibitem [{\citenamefont {Sushko}\ \emph {et~al.}(2013)\citenamefont {Sushko},
  \citenamefont {Bezchastnov}, \citenamefont {Solovʼyov}, \citenamefont
  {Korol}, \citenamefont {Greiner},\ and\ \citenamefont {Solovʼyov}}]{MBNexp}%
  \BibitemOpen
  \bibfield  {author} {\bibinfo {author} {\bibfnamefont {G.~B.}\ \bibnamefont
  {Sushko}}, \bibinfo {author} {\bibfnamefont {V.~G.}\ \bibnamefont
  {Bezchastnov}}, \bibinfo {author} {\bibfnamefont {I.~A.}\ \bibnamefont
  {Solovʼyov}}, \bibinfo {author} {\bibfnamefont {A.~V.}\ \bibnamefont
  {Korol}}, \bibinfo {author} {\bibfnamefont {W.}~\bibnamefont {Greiner}}, \
  and\ \bibinfo {author} {\bibfnamefont {A.~V.}\ \bibnamefont {Solovʼyov}},\
  }\href {http://www.sciencedirect.com/science/article/pii/S0021999113004580}
  {\bibfield  {journal} {\bibinfo  {journal} {Journal of Computational
  Physics}\ }\textbf {\bibinfo {volume} {252}},\ \bibinfo {pages} {404 }
  (\bibinfo {year} {2013})}\BibitemShut {NoStop}%
\bibitem [{\citenamefont {Wistisen}(2014)}]{TobiasUdvikling}%
  \BibitemOpen
  \bibfield  {author} {\bibinfo {author} {\bibfnamefont {T.~N.}\ \bibnamefont
  {Wistisen}},\ }\href {https://link.aps.org/doi/10.1103/PhysRevD.90.125008}
  {\bibfield  {journal} {\bibinfo  {journal} {Phys. Rev. D}\ }\textbf {\bibinfo
  {volume} {90}},\ \bibinfo {pages} {125008} (\bibinfo {year}
  {2014})}\BibitemShut {NoStop}%
\bibitem [{\citenamefont {Lindhard}(1991)}]{Lind91}%
  \BibitemOpen
  \bibfield  {author} {\bibinfo {author} {\bibfnamefont {J.}~\bibnamefont
  {Lindhard}},\ }\href {\doibase 10.1103/PhysRevA.43.6032} {\bibfield
  {journal} {\bibinfo  {journal} {Phys. Rev. A}\ }\textbf {\bibinfo {volume}
  {43}},\ \bibinfo {pages} {6032} (\bibinfo {year} {1991})}\BibitemShut
  {NoStop}%
\bibitem [{\citenamefont {Baier}\ \emph {et~al.}(1998)\citenamefont {Baier},
  \citenamefont {Katkov},\ and\ \citenamefont {Strakhovenko}}]{Baier}%
  \BibitemOpen
  \bibfield  {author} {\bibinfo {author} {\bibfnamefont {V.~N.}\ \bibnamefont
  {Baier}}, \bibinfo {author} {\bibfnamefont {V.~M.}\ \bibnamefont {Katkov}}, \
  and\ \bibinfo {author} {\bibfnamefont {V.~M.}\ \bibnamefont {Strakhovenko}},\
  }\href@noop {} {\emph {\bibinfo {title} {{Electromagnetic processes at high
  energies in oriented single crystals}}}}\ (\bibinfo {year}
  {1998})\BibitemShut {NoStop}%
\bibitem [{\citenamefont {Belkacem}\ \emph {et~al.}(1985)\citenamefont
  {Belkacem}, \citenamefont {Cue},\ and\ \citenamefont
  {Kimball}}]{belkacem_1985}%
  \BibitemOpen
  \bibfield  {author} {\bibinfo {author} {\bibfnamefont {A.}~\bibnamefont
  {Belkacem}}, \bibinfo {author} {\bibfnamefont {N.}~\bibnamefont {Cue}}, \
  and\ \bibinfo {author} {\bibfnamefont {J.}~\bibnamefont {Kimball}},\ }\href
  {\doibase https://doi.org/10.1016/0375-9601(85)90811-4} {\bibfield  {journal}
  {\bibinfo  {journal} {Physics Letters A}\ }\textbf {\bibinfo {volume}
  {111}},\ \bibinfo {pages} {86 } (\bibinfo {year} {1985})}\BibitemShut
  {NoStop}%
\bibitem [{\citenamefont {Kimball}\ \emph {et~al.}(1986)\citenamefont
  {Kimball}, \citenamefont {Cue},\ and\ \citenamefont
  {Belkacem}}]{kimball_1986}%
  \BibitemOpen
  \bibfield  {author} {\bibinfo {author} {\bibfnamefont {J.}~\bibnamefont
  {Kimball}}, \bibinfo {author} {\bibfnamefont {N.}~\bibnamefont {Cue}}, \ and\
  \bibinfo {author} {\bibfnamefont {A.}~\bibnamefont {Belkacem}},\ }\href
  {\doibase https://doi.org/10.1016/0168-583X(86)90461-1} {\bibfield  {journal}
  {\bibinfo  {journal} {Nuclear Instruments and Methods in Physics Research
  Section B: Beam Interactions with Materials and Atoms}\ }\textbf {\bibinfo
  {volume} {13}},\ \bibinfo {pages} {1 } (\bibinfo {year} {1986})}\BibitemShut
  {NoStop}%
\bibitem [{\citenamefont {Andersen}\ \emph {et~al.}(2014)\citenamefont
  {Andersen}, \citenamefont {Andersen}, \citenamefont {Knudsen}, \citenamefont
  {Mikkelsen}, \citenamefont {Thomsen}, \citenamefont {Uggerhøj},
  \citenamefont {Wistisen}, \citenamefont {Esberg}, \citenamefont {Sona},
  \citenamefont {Mangiarotti},\ and\ \citenamefont {Ketel}}]{andersen2014}%
  \BibitemOpen
  \bibfield  {author} {\bibinfo {author} {\bibfnamefont {K.}~\bibnamefont
  {Andersen}}, \bibinfo {author} {\bibfnamefont {S.}~\bibnamefont {Andersen}},
  \bibinfo {author} {\bibfnamefont {H.}~\bibnamefont {Knudsen}}, \bibinfo
  {author} {\bibfnamefont {R.}~\bibnamefont {Mikkelsen}}, \bibinfo {author}
  {\bibfnamefont {H.}~\bibnamefont {Thomsen}}, \bibinfo {author} {\bibfnamefont
  {U.}~\bibnamefont {Uggerhøj}}, \bibinfo {author} {\bibfnamefont
  {T.}~\bibnamefont {Wistisen}}, \bibinfo {author} {\bibfnamefont
  {J.}~\bibnamefont {Esberg}}, \bibinfo {author} {\bibfnamefont
  {P.}~\bibnamefont {Sona}}, \bibinfo {author} {\bibfnamefont {A.}~\bibnamefont
  {Mangiarotti}}, \ and\ \bibinfo {author} {\bibfnamefont {T.}~\bibnamefont
  {Ketel}},\ }\href {\doibase https://doi.org/10.1016/j.physletb.2014.03.055}
  {\bibfield  {journal} {\bibinfo  {journal} {Physics Letters B}\ }\textbf
  {\bibinfo {volume} {732}},\ \bibinfo {pages} {309 } (\bibinfo {year}
  {2014})}\BibitemShut {NoStop}%
\bibitem [{\citenamefont {Andersen}\ \emph {et~al.}(2012)\citenamefont
  {Andersen}, \citenamefont {Andersen}, \citenamefont {Esberg}, \citenamefont
  {Knudsen}, \citenamefont {Mikkelsen}, \citenamefont {Uggerh\o{}j},
  \citenamefont {Sona}, \citenamefont {Mangiarotti}, \citenamefont {Ketel},\
  and\ \citenamefont {Ballestrero}}]{formation}%
  \BibitemOpen
  \bibfield  {author} {\bibinfo {author} {\bibfnamefont {K.~K.}\ \bibnamefont
  {Andersen}}, \bibinfo {author} {\bibfnamefont {S.~L.}\ \bibnamefont
  {Andersen}}, \bibinfo {author} {\bibfnamefont {J.}~\bibnamefont {Esberg}},
  \bibinfo {author} {\bibfnamefont {H.}~\bibnamefont {Knudsen}}, \bibinfo
  {author} {\bibfnamefont {R.}~\bibnamefont {Mikkelsen}}, \bibinfo {author}
  {\bibfnamefont {U.~I.}\ \bibnamefont {Uggerh\o{}j}}, \bibinfo {author}
  {\bibfnamefont {P.}~\bibnamefont {Sona}}, \bibinfo {author} {\bibfnamefont
  {A.}~\bibnamefont {Mangiarotti}}, \bibinfo {author} {\bibfnamefont {T.~J.}\
  \bibnamefont {Ketel}}, \ and\ \bibinfo {author} {\bibfnamefont
  {S.}~\bibnamefont {Ballestrero}} (\bibinfo {collaboration} {CERN NA63}),\
  }\href {\doibase 10.1103/PhysRevLett.108.071802} {\bibfield  {journal}
  {\bibinfo  {journal} {Phys. Rev. Lett.}\ }\textbf {\bibinfo {volume} {108}},\
  \bibinfo {pages} {071802} (\bibinfo {year} {2012})}\BibitemShut {NoStop}%
\bibitem [{\citenamefont {Uggerh\o{}j}(2005)}]{Ulrikcrystals}%
  \BibitemOpen
  \bibfield  {author} {\bibinfo {author} {\bibfnamefont {U.~I.}\ \bibnamefont
  {Uggerh\o{}j}},\ }\href {https://link.aps.org/doi/10.1103/RevModPhys.77.1131}
  {\bibfield  {journal} {\bibinfo  {journal} {Rev. Mod. Phys.}\ }\textbf
  {\bibinfo {volume} {77}},\ \bibinfo {pages} {1131} (\bibinfo {year}
  {2005})}\BibitemShut {NoStop}%
\bibitem [{\citenamefont {Doyle}\ and\ \citenamefont {Turner}(1968)}]{Doyle}%
  \BibitemOpen
  \bibfield  {author} {\bibinfo {author} {\bibfnamefont {P.~A.}\ \bibnamefont
  {Doyle}}\ and\ \bibinfo {author} {\bibfnamefont {P.~S.}\ \bibnamefont
  {Turner}},\ }\href {https://doi.org/10.1107/S0567739468000756} {\bibfield
  {journal} {\bibinfo  {journal} {Acta Crystallographica Section A}\ }\textbf
  {\bibinfo {volume} {24}},\ \bibinfo {pages} {390} (\bibinfo {year}
  {1968})}\BibitemShut {NoStop}%
\bibitem [{\citenamefont {Andersen}\ \emph {et~al.}(1982)\citenamefont
  {Andersen}, \citenamefont {Bonderup}, \citenamefont {Laegsgaard},
  \citenamefont {Marsh},\ and\ \citenamefont {S{\o}rensen}}]{JUAndersen}%
  \BibitemOpen
  \bibfield  {author} {\bibinfo {author} {\bibfnamefont {J.}~\bibnamefont
  {Andersen}}, \bibinfo {author} {\bibfnamefont {E.}~\bibnamefont {Bonderup}},
  \bibinfo {author} {\bibfnamefont {E.}~\bibnamefont {Laegsgaard}}, \bibinfo
  {author} {\bibfnamefont {B.}~\bibnamefont {Marsh}}, \ and\ \bibinfo {author}
  {\bibfnamefont {A.}~\bibnamefont {S{\o}rensen}},\ }\href
  {http://www.sciencedirect.com/science/article/pii/0029554X82905171}
  {\bibfield  {journal} {\bibinfo  {journal} {Nuclear Instruments and Methods
  in Physics Research}\ }\textbf {\bibinfo {volume} {194}},\ \bibinfo {pages}
  {209 } (\bibinfo {year} {1982})}\BibitemShut {NoStop}%
\bibitem [{\citenamefont {M{\o}ller}(1995)}]{SørenPapeBending}%
  \BibitemOpen
  \bibfield  {author} {\bibinfo {author} {\bibfnamefont {S.~P.}\ \bibnamefont
  {M{\o}ller}},\ }\href
  {http://www.sciencedirect.com/science/article/pii/0168900295001816}
  {\bibfield  {journal} {\bibinfo  {journal} {Nuclear Instruments and Methods
  in Physics Research Section A: Accelerators, Spectrometers, Detectors and
  Associated Equipment}\ }\textbf {\bibinfo {volume} {361}},\ \bibinfo {pages}
  {403 } (\bibinfo {year} {1995})}\BibitemShut {NoStop}%
\bibitem [{\citenamefont {Peng}(1999)}]{Peng}%
  \BibitemOpen
  \bibfield  {author} {\bibinfo {author} {\bibfnamefont {L.-M.}\ \bibnamefont
  {Peng}},\ }\href
  {http://www.sciencedirect.com/science/article/pii/S0968432899000335}
  {\bibfield  {journal} {\bibinfo  {journal} {Micron}\ }\textbf {\bibinfo
  {volume} {30}},\ \bibinfo {pages} {625 } (\bibinfo {year}
  {1999})}\BibitemShut {NoStop}%
\bibitem [{\citenamefont {Babaev}\ and\ \citenamefont
  {Dabagov}(2012)}]{ElectronScat}%
  \BibitemOpen
  \bibfield  {author} {\bibinfo {author} {\bibfnamefont {A.}~\bibnamefont
  {Babaev}}\ and\ \bibinfo {author} {\bibfnamefont {S.}~\bibnamefont
  {Dabagov}},\ }\href {\doibase 10.1140/epjp/i2012-12062-6} {\bibfield
  {journal} {\bibinfo  {journal} {The European Physical Journal Plus}\ }\textbf
  {\bibinfo {volume} {127}} (\bibinfo {year} {2012}),\
  10.1140/epjp/i2012-12062-6}\BibitemShut {NoStop}%
\bibitem [{\citenamefont {Tanabashi}\ \emph {et~al.}(2018)\citenamefont
  {Tanabashi} \emph {et~al.}}]{PDG}%
  \BibitemOpen
  \bibfield  {author} {\bibinfo {author} {\bibfnamefont {M.}~\bibnamefont
  {Tanabashi}} \emph {et~al.} (\bibinfo {collaboration} {Particle Data
  Group}),\ }\href {https://link.aps.org/doi/10.1103/PhysRevD.98.030001}
  {\bibfield  {journal} {\bibinfo  {journal} {Phys. Rev. D}\ }\textbf {\bibinfo
  {volume} {98}},\ \bibinfo {pages} {030001} (\bibinfo {year}
  {2018})}\BibitemShut {NoStop}%
\bibitem [{\citenamefont {Kitagawa}\ and\ \citenamefont
  {Ohtsuki}(1973)}]{Scattering}%
  \BibitemOpen
  \bibfield  {author} {\bibinfo {author} {\bibfnamefont {M.}~\bibnamefont
  {Kitagawa}}\ and\ \bibinfo {author} {\bibfnamefont {Y.}~\bibnamefont
  {Ohtsuki}},\ }\href {\doibase 10.1103/PhysRevB.8.3117} {\bibfield  {journal}
  {\bibinfo  {journal} {Phys. Rev. B}\ }\textbf {\bibinfo {volume} {8}}
  (\bibinfo {year} {1973}),\ 10.1103/PhysRevB.8.3117}\BibitemShut {NoStop}%
\bibitem [{\citenamefont {Jensen}\ and\ \citenamefont
  {S\o{}rensen}(2013)}]{Elektrondensity}%
  \BibitemOpen
  \bibfield  {author} {\bibinfo {author} {\bibfnamefont {T.~V.}\ \bibnamefont
  {Jensen}}\ and\ \bibinfo {author} {\bibfnamefont {A.~H.}\ \bibnamefont
  {S\o{}rensen}},\ }\href {\doibase 10.1103/PhysRevA.87.022902} {\bibfield
  {journal} {\bibinfo  {journal} {Phys. Rev. A}\ }\textbf {\bibinfo {volume}
  {87}},\ \bibinfo {pages} {022902} (\bibinfo {year} {2013})}\BibitemShut
  {NoStop}%
\bibitem [{\citenamefont {Landau}\ and\ \citenamefont
  {Lifshitz}(1979)}]{Landau}%
  \BibitemOpen
  \bibfield  {author} {\bibinfo {author} {\bibfnamefont {L.}~\bibnamefont
  {Landau}}\ and\ \bibinfo {author} {\bibfnamefont {E.}~\bibnamefont
  {Lifshitz}},\ }\href@noop {} {\emph {\bibinfo {title} {The Classical Theory
  of Fields}}},\ \bibinfo {edition} {4th}\ ed.\ (\bibinfo  {publisher}
  {Pergamon press},\ \bibinfo {year} {1979})\BibitemShut {NoStop}%
\bibitem [{\citenamefont {Wistisen}(2015)}]{TobiasSpredning}%
  \BibitemOpen
  \bibfield  {author} {\bibinfo {author} {\bibfnamefont {T.~N.}\ \bibnamefont
  {Wistisen}},\ }\href {\doibase 10.1103/PhysRevD.92.045045} {\bibfield
  {journal} {\bibinfo  {journal} {Phys. Rev. D}\ }\textbf {\bibinfo {volume}
  {92}},\ \bibinfo {pages} {045045} (\bibinfo {year} {2015})}\BibitemShut
  {NoStop}%
\bibitem [{Note2()}]{Note2}%
  \BibitemOpen
  \bibinfo {note}
  {Https://gitlab.au.dk/au483748/channelingradiation.git}\BibitemShut {NoStop}%
\bibitem [{\citenamefont {Hofmann}(2004)}]{undulator}%
  \BibitemOpen
  \bibfield  {author} {\bibinfo {author} {\bibfnamefont {A.}~\bibnamefont
  {Hofmann}},\ }\href@noop {} {\emph {\bibinfo {title} {The Physics of
  Synchrotron Radiation}}},\ \bibinfo {edition} {1st}\ ed.\ (\bibinfo
  {publisher} {Cambridge university press},\ \bibinfo {year} {2004})\
  Chap.~\bibinfo {chapter} {8}\BibitemShut {NoStop}%
\bibitem [{\citenamefont {Palazzi}(1968)}]{coherentbremstrahlung}%
  \BibitemOpen
  \bibfield  {author} {\bibinfo {author} {\bibfnamefont {G.~D.}\ \bibnamefont
  {Palazzi}},\ }\href {https://link.aps.org/doi/10.1103/RevModPhys.40.611}
  {\bibfield  {journal} {\bibinfo  {journal} {Rev. Mod. Phys.}\ }\textbf
  {\bibinfo {volume} {40}},\ \bibinfo {pages} {611} (\bibinfo {year}
  {1968})}\BibitemShut {NoStop}%
\bibitem [{\citenamefont {Ter-Mikaelian}(1972)}]{coherentbremstrahlung2}%
  \BibitemOpen
  \bibfield  {author} {\bibinfo {author} {\bibfnamefont {M.~L.}\ \bibnamefont
  {Ter-Mikaelian}},\ }\href@noop {} {\emph {\bibinfo {title} {High-energy
  electromagnetic processes in condensed media}}}\ (\bibinfo  {publisher}
  {Wiley-Interscience},\ \bibinfo {year} {1972})\BibitemShut {NoStop}%
\bibitem [{\citenamefont {Bak}\ \emph {et~al.}(1985)\citenamefont {Bak},
  \citenamefont {Ellison}, \citenamefont {Marsh}, \citenamefont {Meyer},
  \citenamefont {Pedersen}, \citenamefont {Petersen}, \citenamefont
  {Uggerh{\o}j}, \citenamefont {{\o}stergaard}, \citenamefont {M{\o}ller},
  \citenamefont {S{\o}rensen},\ and\ \citenamefont {Suffert}}]{Baketal}%
  \BibitemOpen
  \bibfield  {author} {\bibinfo {author} {\bibfnamefont {J.}~\bibnamefont
  {Bak}}, \bibinfo {author} {\bibfnamefont {J.}~\bibnamefont {Ellison}},
  \bibinfo {author} {\bibfnamefont {B.}~\bibnamefont {Marsh}}, \bibinfo
  {author} {\bibfnamefont {F.}~\bibnamefont {Meyer}}, \bibinfo {author}
  {\bibfnamefont {O.}~\bibnamefont {Pedersen}}, \bibinfo {author}
  {\bibfnamefont {J.}~\bibnamefont {Petersen}}, \bibinfo {author}
  {\bibfnamefont {E.}~\bibnamefont {Uggerh{\o}j}}, \bibinfo {author}
  {\bibfnamefont {K.}~\bibnamefont {{\o}stergaard}}, \bibinfo {author}
  {\bibfnamefont {S.}~\bibnamefont {M{\o}ller}}, \bibinfo {author}
  {\bibfnamefont {A.}~\bibnamefont {S{\o}rensen}}, \ and\ \bibinfo {author}
  {\bibfnamefont {M.}~\bibnamefont {Suffert}},\ }\href
  {http://www.sciencedirect.com/science/article/pii/0550321385902305}
  {\bibfield  {journal} {\bibinfo  {journal} {Nuclear Physics B}\ }\textbf
  {\bibinfo {volume} {254}},\ \bibinfo {pages} {491 } (\bibinfo {year}
  {1985})}\BibitemShut {NoStop}%
\bibitem [{108()}]{1080ti}%
  \BibitemOpen
  \href@noop {} {\emph {\bibinfo {title} {White Paper, NVIDIA GeForce GTX 1080,
  Gaming Perfected}}},\ \bibinfo {organization} {Nvidia}\BibitemShut {NoStop}%
\end{thebibliography}%

\end{document}